\documentclass[english]{article}
\usepackage[T1]{fontenc}
\usepackage[latin9]{inputenc}
\usepackage{geometry}
\usepackage{amsmath}
\usepackage{amssymb}
\usepackage{graphicx}
\usepackage{cite}

\makeatletter

\newcommand\rmd{\mathrm{d}}
\newcommand\erfc{\mathrm{erfc}}

\makeatother

\begin{document}
\title{Approximation of the first passage time distribution for the birth--death
processes}
\author{Aleksejus Kononovicius, Vygintas Gontis}
\date{Institute of Theoretical Physics and Astronomy, Vilnius University}
\maketitle
\begin{abstract}
We propose a general method to obtain approximation of the first passage
time distribution for the birth--death processes. We rely on the
general properties of birth--death processes, Keilson's theorem and
the concept of Riemann sum to obtain closed--form expressions. We
apply the method to the three selected birth--death processes and
the sophisticated order--book model exhibiting long--range memory.
We discuss how our approach contributes to the competition between
spurious and true long--range memory models.
\end{abstract}

\section{Introduction}

Markov chains, and more specifically birth--death processes, are
of great interest in the modeling of biological and socio--economic
systems \cite{Lloyd2001Science,Gonzalez2008Nature,Sundarapandian2009PHILearning}.
While usually birth--death processes converge to some fixed point,
in which birth and death rates are approximately equal, there are
also a few examples which do not converge and system--wide fluctuations
persist \cite{Alfarano2009Dyncon,Flache2011JCR,Kononovicius2014EPJB}.
This strand of research is important and present in various fields
of science, because it could provide an answer to Axelrod's question
about the persistence of diversity in social systems \cite{Axelrod1997JConfRes,Flache2017JASSS}.
The question of diversity, and especially the collapses of diversity,
is also important to the financial markets. Some recent financial
agent--based models (abbr. ABMs) have demonstrated that power--law
distributions emerge, when the diversity breaks down and agents start
to behave similarly \cite{Kononovicius2012PhysA,Kaizoji2015,Cocco2017JEIC,Biondo2018SEF,Franke2018,Llacay2018CompMatOrgT}.
One of the approaches, based on the birth--death process, \cite{Kononovicius2012PhysA}
has explicitly shown that the long--range memory phenomenon could
emerge due to the same underlying reasons.

Long--range memory phenomenon can be reproduced using non--linear
Markov processes \cite{Kaulakys2015MPLB,Kononovicius2015PhysA} and
models with embedded memory, such as ones built using fractional Brownian
motion \cite{Ding1995fbm,Dieker2003fbm,Bassler2006PhysA,McCauley2007PhysA,McCauley2008PhysA},
CTRW framework \cite{Kasprzak2010EPJB,Hilfer2015A,Denys2016APPA,Caceres2017EPJB,Gubiec2017EPJB,Kutner2017EPJB}
or ARCH framework \cite{Giraitis2009,Giraitis2012,Tayefi2012}. Though
these models are able to reproduce few similar statistical features,
they differ in their first passage times \cite{Ding1995fbm,Redner2001Cambridge,Metzler2014WorldScientific}.
It is known that first passage time probability density function (PDF)
of fractional Brownian motion have a region with power--law dependence
with exponent $H-2$ \cite{Ding1995fbm}, while first passage time
PDF of one--dimensional Markov processes has power--law exponent
of $-3/2$ \cite{Redner2001Cambridge,Metzler2014WorldScientific}.
Which naturally suggests a method to determine whether the observed
long--range memory phenomenon is spurious or not.

The problem arises, because the current knowledge describes the asymptotic
behavior of the first passage time PDFs in very broad terms. There
is an alternative description using Laplace transforms, yet these
are often expressed using infinite sums and special functions \cite{Borodin2012Birkhauser}.
With a few notable exceptions, usually such Laplace transforms are
not invertible and even if they are invertible they might not have
explicit closed--forms. Consequently there has been numerical efforts
to approximate the inverses \cite{Abate2006INFORMS}.

Here we propose a general method to analytically approximate specific
first passage times of birth--death processes. We build upon our
previous work \cite{Gontis2012ACS} in which we have obtained an approximation
for a specific first passage time, called inter--burst duration,
PDF of the continuous Bessel process. We have shown that the approximation
could be applied to a diffusion process transformable into Bessel
process (with stochastic process described in \cite{Ruseckas2010PhysRevE}
provided as an example). Notably the approximation had a divergence
problem for short times, which we now solve by proposing to study
birth--death processes with finite and discrete state space.

The paper is organized as follows. In Section~\ref{sec:burstCont}
we introduce the concept of the bursting statistics in the context
of the continuous Bessel process. In Section~\ref{sec:approximation}
we introduce approximation method based on Keilson's theorem \cite{Keilson1979AMS}
and the results reported in \cite{Gontis2012ACS}. In Section~\ref{sec:burstBD}
we approximate bursting statistics of a few selected birth--death
processes: Bessel--like, Ornstein--Uhlenbeck and imitation processes.
In Section~\ref{sec:sophisticated-model} we demonstrate that this
approach can be used to fit the bursting statistics of a sophisticated
model. Concluding remarks are provided in Section~\ref{sec:conclusions}.

\section{Bursting statistics of the continuous Bessel process\label{sec:burstCont}}

\begin{figure}[!h]
\begin{centering}
\includegraphics[width=0.4\textwidth]{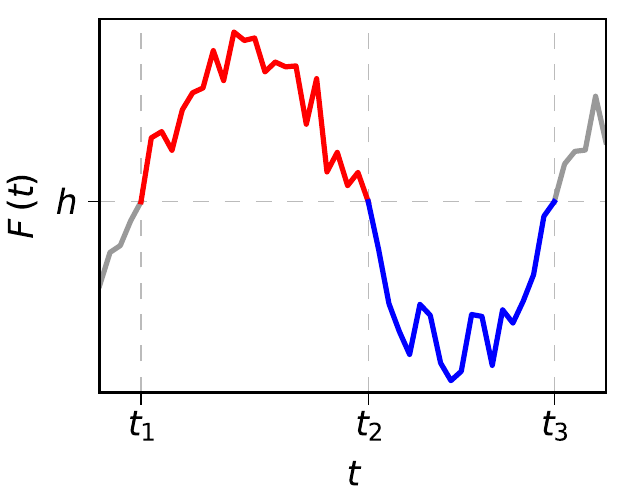}
\par\end{centering}
\caption{(color online) Sample time series (gray curve) with one time period
fully spent above the threshold (highlighted in red) and one time
period fully spent below the threshold (highlighted in blue). Duration
$t_{2}-t_{1}$ is by definition a sample of burst duration $\tau_{h}$,
while duration $t_{3}-t_{2}$ is by definition a sample of inter--burst
duration $\theta_{h}$.\label{fig:sample-definitions}}
\end{figure}

In \cite{Gontis2012ACS,Gontis2017PhysA,Gontis2017Entropy,Gontis2018PhysA}
we have considered burst durations and inter--burst durations of
empirical time series and time series generated by stochastic processes.
We have defined burst duration $\tau_{h}$ as a time spent above the
certain threshold $h$:
\begin{equation}
\tau_{h}=\mathrm{Inf}\left\{ t>0:F\left(t\right)\leq h|F\left(0\right)=h+\epsilon\right\} ,
\end{equation}
here $F\left(t\right)$ describes time evolution of a stochastic process
or an empirical time series, while $\epsilon$ is an infinitesimally
small positive number. In a similar fashion we have defined inter--burst
duration $\theta_{h}$ as a time spent below the threshold:
\begin{equation}
\theta_{h}=\mathrm{Inf}\left\{ t>0:F\left(t\right)\geq h|F\left(0\right)=h-\epsilon\right\} .
\end{equation}
In Fig.~\ref{fig:sample-definitions} we have illustrated these concepts
on a sample time series. This sample contains a single example of
a burst duration and a single example of an inter--burst duration.

In \cite{Gontis2012ACS} we have obtained an approximation for the
PDF of $\theta_{h}$ for the continuous Bessel process. This was possible,
because first passage times from $y_{0}$ to $h$ (with $0<y_{0}<h$)
for the continuous Bessel process are known \cite{Borodin2012Birkhauser}:
\begin{equation}
p_{y_{0},h}^{\left(\nu\right)}\left(\theta_{h}\right)=\frac{h^{\nu-2}}{y_{0}^{\nu}}\sum_{k=1}^{\infty}\frac{j_{\nu,k}J_{\nu}\left(\frac{y_{0}}{h}j_{\nu,k}\right)}{J_{\nu+1}\left(j_{\nu,k}\right)}\exp\left(-\frac{j_{\nu,k}^{2}}{2h^{2}}\theta_{h}\right),\label{eq:bessPdfSum}
\end{equation}
here $\nu$ is the index of the continuous Bessel process, $J_{\nu}\left(x\right)$
is a Bessel function of a first kind and $j_{\nu,k}$ is $k$-th zero
of $J_{\nu}\left(x\right)$.

\begin{figure}[!h]
\begin{centering}
\includegraphics[width=0.4\textwidth]{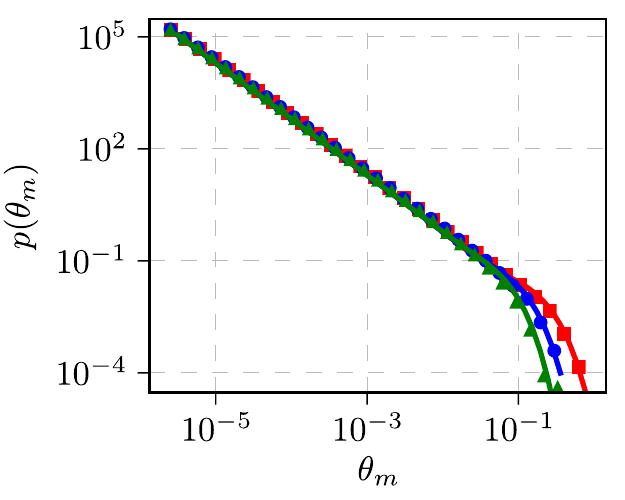}
\par\end{centering}
\caption{(color online) Inter--burst duration PDF of the continuous Bessel
process with three distinct indices, $\nu=0.5$ (red squares), $1.5$
(blue circles) and $2.5$ (green triangles), at level $h=0.7$. Numerical
results are approximated by Eq.~(\ref{eq:bessPdfInt}) (respectively
colored curves).\label{fig:contBessIB}}
\end{figure}

Let $y_{0}=h-\epsilon$, where $\epsilon$ is an infinitesimally small
positive number, then Eq.~(\ref{eq:bessPdfSum}) can be approximated
by an integral:
\begin{align}
p\left(\theta_{h}\right) & \approx C_{1}\sum_{k=1}^{\infty}j_{\nu,k}^{2}\exp\left(-\frac{j_{\nu,k}^{2}}{2h_{y}^{2}}\theta_{h}\right)\approx C_{2}\int_{j_{\nu,1}}^{\infty}x^{2}\exp\left(-\frac{x^{2}}{2h_{y}^{2}}\theta_{h}\right)\rmd x=\nonumber \\
 & =C_{2}\left[\frac{h_{y}^{2}j_{\nu,1}\exp\left(-\frac{j_{\nu,1}^{2}}{2h_{y}^{2}}\theta_{h}\right)}{\theta_{h}}+\sqrt{\frac{\pi}{2}}\frac{h_{y}^{3}\erfc\left(\frac{j_{\nu,1}}{\sqrt{2}h_{y}}\sqrt{\theta_{h}}\right)}{\theta_{h}^{3/2}}\right].\label{eq:bessPdfInt}
\end{align}
This approximation is possible only because $j_{\nu,k}$ grow almost
linearly with $k$. For various $\nu$ only the first few values deviate
from this tendency. In this case we can treat the sum as if it was
a Riemann sum and replace the sum by integral. As we can see in Fig.~\ref{fig:contBessIB}
this approximation works rather well for the continuous Bessel process.

The normalization of this approximation diverges unless some minimum
value of $\theta_{h}$ is assumed. To keep the expression general
we have left the normalization constants denoted by $C_{i}$. While
the minimum value of $\theta_{h}$ can be easily justified from empirical
point of view, assuming equivalence to the discretization period,
such assumption isn't as transparent in case of a continuous model.

Note that the asymptotic behavior of Eq.~(\ref{eq:bessPdfInt}) is
consistent with what would be expected of a one--dimensional Markov
processes \cite{Redner2001Cambridge}. For the small $\theta_{h}$the
second term is the largest and therefore power lay decay is observed
with exponent $-3/2$. For the larger $\theta_{h}$ the first term
becomes the largest and the exponential cutoff is observed.

It is not possible to obtain approximation for $\tau_{m}$ of the
continuous Bessel process, because for the most $\nu$ the Bessel
process escapes towards infinity and therefore does not always hit
$h$. Consequently there is no general formula for the hitting times
of the Bessel process for $0<h<y_{0}$ case \cite{Borodin2012Birkhauser}.

It is important to note that Eq.~(\ref{eq:bessPdfInt}) applies to
any process, which can be transformed into Bessel process by means
of Lamperti transformation. Though if the Lamperti transformation
stipulates inverse relationship between processes, then Eq.~(\ref{eq:bessPdfInt})
would approximate burst duration PDF of that process instead. In \cite{Gontis2012ACS}
we have considered one such case. We have transformed a non--linear
stochastic process exhibiting long--range memory \cite{Ruseckas2011PhysRevE}
into Bessel process by means of Lamperti transformation. We have shown
that Eq.~(\ref{eq:bessPdfInt}) provides a good fit for the burst
duration PDF of numerical simulations of the non--linear stochastic
process exhibiting spurious long--range memory.

\section{Approximation of inter--burst duration for any birth--death process\label{sec:approximation}}

Through Keilson's theorem \cite{Keilson1979AMS,Gong2012JTP} it is
well known that Laplace transform of a first passage time from state
$k$ to state $n$, with $0<k<n$, is given by:
\begin{equation}
\left\langle e^{-sT_{k,n}}\right\rangle =\frac{\prod{}_{i=1}^{n}\frac{\lambda_{i}^{\left(n\right)}}{s+\lambda_{i}^{\left(n\right)}}}{\prod{}_{i=1}^{k}\frac{\lambda_{i}^{\left(k\right)}}{s+\lambda_{i}^{\left(k\right)}}},\quad s\ge0.
\end{equation}
In the above $\lambda_{i}^{\left(n\right)}$ are the sorted positive
eigenvalues of negative birth--death process generator matrix truncated
at rank $n$. Inverse Laplace transforms of this expression is not
feasible for many reasonable values of $n$ and $k$. Yet we have
observed that for some small $n$ and $k=n-1$ inverse Laplace transform
yields a sum of exponents whose rates are given by $\lambda_{i}^{\left(n\right)}$.
As this sum has a similar form to the one in Eq.~(\ref{eq:bessPdfInt})
we propose to apply the same approximation method:
\begin{align}
p\left(\theta_{h}\right) & \approx\frac{1}{\sqrt{\lambda_{n}^{\left(n\right)}}-\sqrt{\lambda_{1}^{\left(n\right)}}}\int{}_{\sqrt{\lambda_{1}^{\left(n\right)}}}^{\sqrt{\lambda_{n}^{\left(n\right)}}}x^{2}\exp\left(-x^{2}\theta_{h}\right)\rmd x=\nonumber \\
 & =\frac{1}{4\left(\sqrt{\lambda_{n}^{\left(n\right)}}-\sqrt{\lambda_{1}^{\left(n\right)}}\right)}\left[2\frac{\sqrt{\lambda_{1}^{\left(n\right)}}\exp\left(-\lambda_{1}^{\left(n\right)}\theta_{h}\right)-\sqrt{\lambda_{n}^{\left(n\right)}}\exp\left(-\lambda_{n}^{\left(n\right)}\theta_{h}\right)}{\theta_{h}}+\right.\label{eq:bdApproxInt}\\
 & \left.+\sqrt{\pi}\frac{\erfc\left(\sqrt{\lambda_{1}^{\left(n\right)}\theta_{h}}\right)-\erfc\left(\sqrt{\lambda_{n}^{\left(n\right)}\theta_{h}}\right)}{\theta_{h}^{3/2}}\right]=I\left(\theta_{h},\lambda_{1}^{\left(n\right)},\lambda_{n}^{\left(n\right)}\right),\nonumber 
\end{align}
here $\lambda_{1}^{\left(n\right)}$ is the first (smallest) positive
eigenvalue, $\lambda_{n}^{\left(n\right)}$ is the last (largest)
positive eigenvalue and we introduce $I\left(\theta_{h},\lambda_{1}^{\left(n\right)},\lambda_{n}^{\left(n\right)}\right)$
to simplify further notation. Similar truncation idea was used in
\cite{Dassios2018FPT} to approximate asymptotic behavior of diffusion
processes. This approximation will be valid as long as $\sqrt{\lambda_{i}^{\left(n\right)}}$
grows linearly in wide range of $i$. It is worth to note that this
also seems to be the only way to obtain $p\left(\theta_{h}\right)\propto\theta_{h}^{-3/2}$,
which is a general result for any one--dimensional Markov process.
Therefore it seems that every birth--death process should have a
reasonably wide region in which $\sqrt{\lambda_{i}^{\left(n\right)}}$
grow linearly.

Note that the first order approximation, Eq.~(\ref{eq:bdApproxInt}),
is usually insufficient to approximate the PDF of $\theta_{h}$, because
usually the impact of $\lambda_{1}$ term becomes undervalued. To
improve the overall approximation we propose to use integral approximation
from the second term in the sum onwards and to account for the first
term separately. Then the second order approximation of the inter--burst
duration PDF would have the following form:
\begin{equation}
p\left(\theta_{h}\right)\approx\rho\lambda_{1}^{\left(n\right)}\exp\left[-\lambda_{1}^{\left(n\right)}\theta_{h}\right]+\left(1-\rho\right)I\left(\theta_{h},\lambda_{2}^{\left(n\right)},\lambda_{n}^{\left(n\right)}\right).\label{eq:bdApprox2}
\end{equation}
Here $\rho$ is used to estimate the relative impact of the first
eigenvalue term. It is evaluated by comparing the contributions to
the resulting first moments, $\bar{\theta}_{t}$, of the first exponent
(first term) and of the PDF approximated by the integral (second term).
Then the exact first moment of the respective hitting time of a birth--death
process ($\bar{\theta}_{h}$ on the right hand side of the equation):
\begin{equation}
\bar{\theta}_{h}=\rho\cdot\frac{1}{\lambda_{1}^{\left(n\right)}}+\left(1-\rho\right)\cdot\frac{1}{\sqrt{\lambda_{2}^{\left(n\right)}\lambda_{n}^{\left(n\right)}}}\quad\Rightarrow\quad\rho=\frac{\lambda_{1}\left(1-\sqrt{\lambda_{2}^{\left(n\right)}\lambda_{n}^{\left(n\right)}}\bar{\theta}_{h}\right)}{\lambda_{1}-\sqrt{\lambda_{2}^{\left(n\right)}\lambda_{n}^{\left(n\right)}}}.
\end{equation}
Note that $\bar{\theta}_{h}$ as well as the other moments of hitting
times for any birth--death process is known explicitly and can be
obtained exactly (see \cite{Jouini2008MatMetOpRes} for more details).
Thus if there is a need to further improve the approximation, then
one could split more terms from the sum and estimate their relative
impact based on the higher moments. Here we limit ourselves just to
the first moment as this proves sufficient in the examples we have
selected.

\section{Bursting statistics of the selected birth--death processes\label{sec:burstBD}}

In this section we will explore bursting statistics of a few selected
birth--death processes. We will consider Bessel--like, Ornstein--Uhlenbeck
and imitation processes. Using these three examples we will show that
the proposed approximation method works rather well.

While these three examples are rather different, they share one property
-- their states are equivalent in statistical sense:
\begin{equation}
X\overset{s.s.}{\equiv}N-X.\label{eq:stateEquiv}
\end{equation}
This property means that the burst duration and the inter--burst
duration are also equivalent in statistical sense:
\begin{equation}
\theta_{h}\overset{s.s.}{\equiv}\tau_{1-h}.\label{eq:burstEquiv}
\end{equation}
Thus knowing the PDF of $\theta_{h}$ also gives us a PDF for $\tau_{1-h}$.

\subsection{The birth--death Bessel--like process}

Let us define the transition rates for the birth--death Bessel--like
process:
\begin{equation}
\lambda_{b}^{+}=\frac{N^{2}}{2}\left(1+\frac{\nu_{1}+\frac{1}{2}}{X}-\frac{\nu_{2}+\frac{1}{2}}{N-X}\right),\quad\lambda_{b}^{-}=\frac{N^{2}}{2}\left(1-\frac{\nu_{1}+\frac{1}{2}}{X}+\frac{\nu_{2}+\frac{1}{2}}{N-X}\right),\label{eq:symBesselRates}
\end{equation}
here $N$ is the number of the available states, $X$ is the current
state, $\nu_{1}$ and $\nu_{2}$ are the indices of the process. To
keep these rates well behaved let us require that $\nu_{i}=\frac{1}{2}+n$
with $n\in\mathbb{N}_{0}$. With $\nu_{1}>-\frac{1}{2}$ and $\nu_{2}>-\frac{1}{2}$,
in the continuous limit $N\rightarrow\infty$ (defining $x=X/N$),
we get a stochastic process trapped between $x=0$ and $x=1$:
\begin{equation}
\rmd x=\left(\nu+\frac{1}{2}\right)\frac{2x-1}{x(1-x)}\rmd t+\rmd W.
\end{equation}

Two special cases with unequal indices deserve out attention. In the
continuous limit with $\nu_{1}=\nu$ and $\nu_{2}=-\frac{1}{2}$ we
recover SDE of the continuous Bessel process:
\begin{equation}
\rmd x=\left(\nu+\frac{1}{2}\right)\frac{\rmd t}{x}+\rmd W.
\end{equation}
Similar birth--death process was proposed in \cite{Kounta2017} and
shown to asymptotically approach the continuous Bessel process. While
with $\nu_{1}=-\frac{1}{2}$ and $\nu_{2}=\nu$ we get a stochastic
process, whose potential is a shifted reflection of the Bessel process
potential. In this case the potential is defined for $x\in\left(-\infty;1\right)$
instead of $\left(0;+\infty\right)$ as for the Bessel process. The
corresponding SDE is given by:
\begin{equation}
\rmd x=-\left(\nu+\frac{1}{2}\right)\frac{\rmd t}{1-x}+\rmd W.
\end{equation}
Here, seeking to retain equivalence properties Eqs.~(\ref{eq:stateEquiv})
and (\ref{eq:burstEquiv}), we will consider only the symmetric version
of this birth--death process assuming that $\nu_{1}=\nu_{2}=\nu$.

\begin{figure}
\begin{centering}
\includegraphics[width=0.7\textwidth]{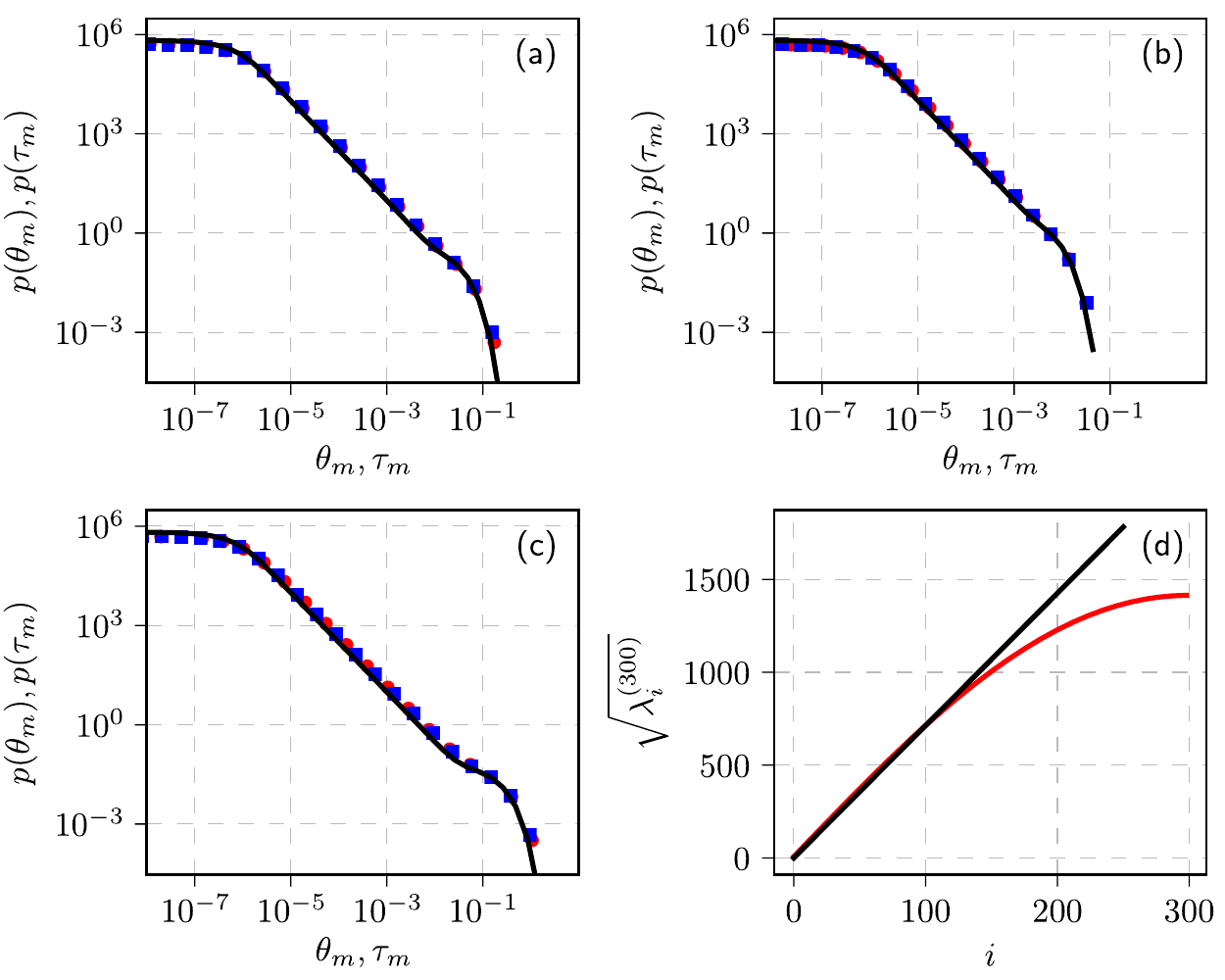}
\par\end{centering}
\caption{(color online) Burst and inter--burst duration PDFs ((a), (b) and
(c)) for the symmetric case of birth--death Bessel--like process
and square root of its eigenvalue spectrum (sub--figure (d)). In
(a), (b) and (c) red circles represent numerical inter--burst duration
PDF, blue squares represent numerical burst duration PDF, while black
curve was obtained by evaluating Eq.~(\ref{eq:bdApprox2}). In (d)
red curve represents a square roots of the eigenvalues, while black
curve show the best linear fit. Parameters of the process were set
as follows: (a) $\nu=0.5$, $N=10^{3}$, $h=0.3$ (for inter--burst
duration) and $0.7$ (for burst duration), (b) $\nu=1.5$, $N=10^{3}$,
$h=0.2$ (for inter--burst duration) and $0.8$ (for burst duration),
(c) $\nu=2.5$, $N=10^{3}$, $h=0.7$ (for inter--burst duration)
and $0.3$ (for burst duration), (d) was obtained with same parameters
as (a).\label{fig:symBessel}}
\end{figure}

In Fig.~\ref{fig:symBessel} we show that burst and inter--burst
durations in the symmetric version of this model are indeed equivalent
in statistical sense, as blue squares and red circles trace the same
shape. Furthermore the approximation by Eq.~(\ref{eq:bdApprox2})
provides rather good approximation for that shape. Also in Fig.~\ref{fig:symBessel}
(d) we show that the square root of the eigenvalue spectrum indeed
grows linearly in a certain region.

\subsection{The birth--death Ornstein--Uhlenbeck process}

Let us define the transition rates for the birth--death Ornstein--Uhlenbeck
process:
\begin{equation}
\lambda_{ou}^{+}=N^{2}\left(1-\frac{X}{N}\right),\quad\lambda_{ou}^{-}=N^{2}\cdot\frac{X}{N}.\label{eq:ouRates}
\end{equation}
For these rates the corresponding SDE is given by:
\begin{equation}
\rmd x=2N\left(\frac{1}{2}-x\right)\rmd t+\rmd W,
\end{equation}
which describes an Ornstein--Uhlenbeck process with the relaxation
rate $\theta=2N$ to the resting point $\mu=1/2$. Due to the form
of these rates this process is defined for $x\in\left[0,1\right]$
(alternatively $X\in\left[0,N\right]$).

\begin{figure}
\begin{centering}
\includegraphics[width=0.7\textwidth]{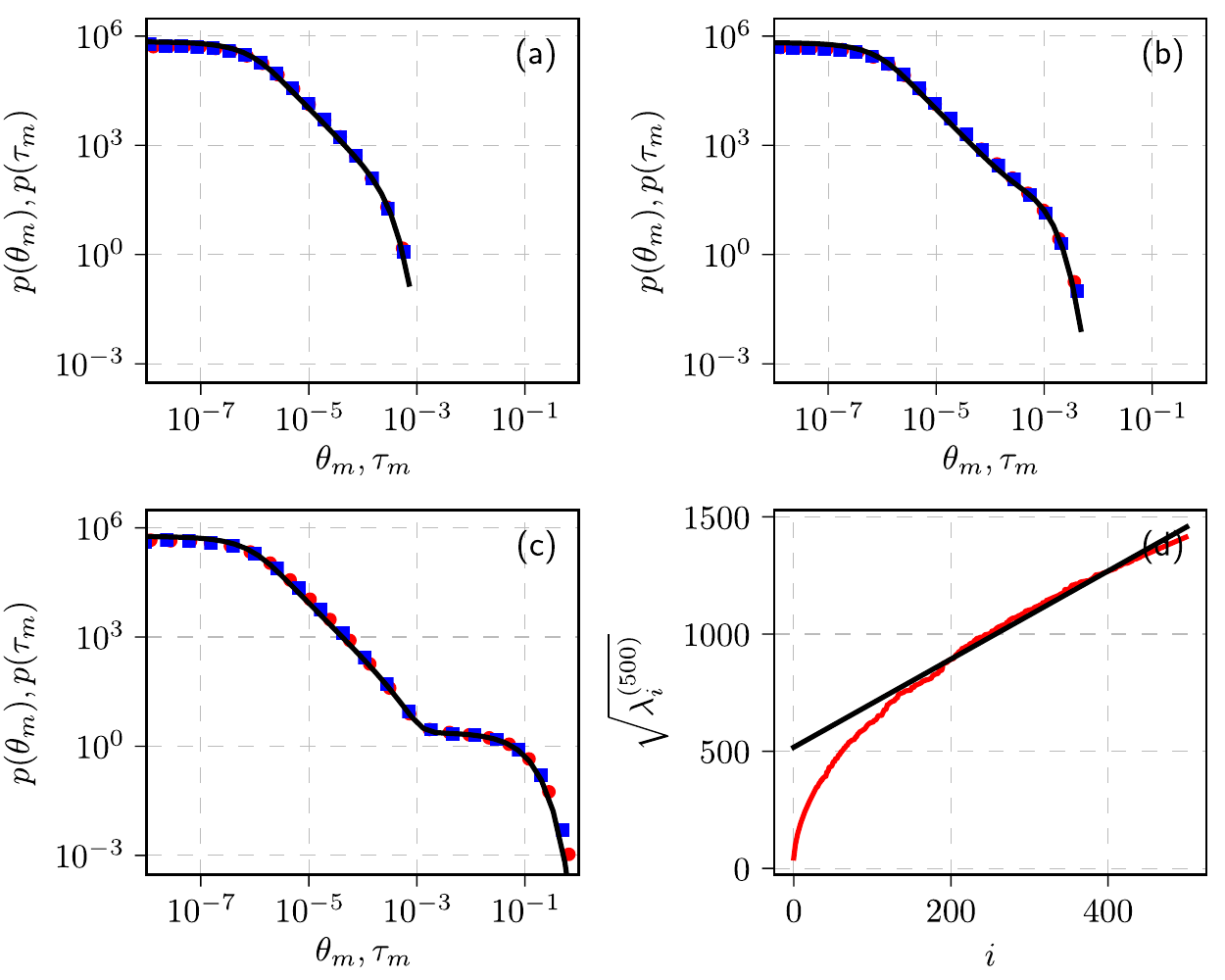}
\par\end{centering}
\caption{(color online) Burst and inter--burst duration PDFs ((a), (b) and
(c)) for the Ornstein--Uhlenbeck process and square root of its eigenvalue
spectrum (sub--figure (d)). In (a), (b) and (c) red circles represent
numerical inter--burst duration PDF, blue squares represent numerical
burst duration PDF, while black curve was obtained by evaluating Eq.~(\ref{eq:bdApprox2}).
In (d) red curve represents a square roots of the eigenvalues, while
black curve show the best linear fit. Parameters of the process were
set as follows: (a) $N=10^{3}$, $h=0.45$ (for inter--burst duration)
and $0.55$ (for burst duration), (b) $\varepsilon=1$, $N=10^{3}$,
$h=0.5$ (for inter--burst and burst duration), (c) $N=10^{3}$,
$h=0.55$ (for inter--burst duration) and $0.45$ (for burst duration),
(d) was obtained with same parameters as (b).\label{fig:ou}}
\end{figure}

In Fig.~\ref{fig:ou} we show that burst and inter--burst durations
in this model are indeed equivalent in statistical sense, as blue
squares and red circles trace the same shape. Furthermore the approximation
by Eq.~(\ref{eq:bdApprox2}) provides rather good approximation for
that shape. Also in Fig.~\ref{fig:ou} (d) we show that the square
root of the eigenvalue spectrum indeed grows linearly in a certain
region.

\subsection{The imitation process}

Another birth--death process we consider in this paper is inspired
by Kirman's model \cite{Kirman1993QJE}. This process is of particular
interest, because it is known to exhibit long--range memory when
applied to model financial markets \cite{Kononovicius2012PhysA}.
The rates of this imitation process are given by:
\begin{equation}
\lambda_{k}^{+}=\left(N-X\right)\left(\varepsilon+X\right),\quad\lambda_{k}^{-}=X\left[\varepsilon+\left(N-X\right)\right].\label{eq:kirmanRates}
\end{equation}
This process can be interpreted as follows. $N$ agents switch between
two states independently with rate $\varepsilon$ and also due to
pairwise interactions, which happen at unit rate. Note that here we
use simplified rates, while a more sophisticated forms can be also
assumed to give the process a richer behavior \cite{Kononovicius2012PhysA}.

In continuous limit from these rates we recover the following SDE:
\begin{equation}
\rmd x=\varepsilon\left(1-2x\right)\rmd t+\sqrt{2x(1-x)}\rmd W.
\end{equation}

\begin{figure}
\begin{centering}
\includegraphics[width=0.7\textwidth]{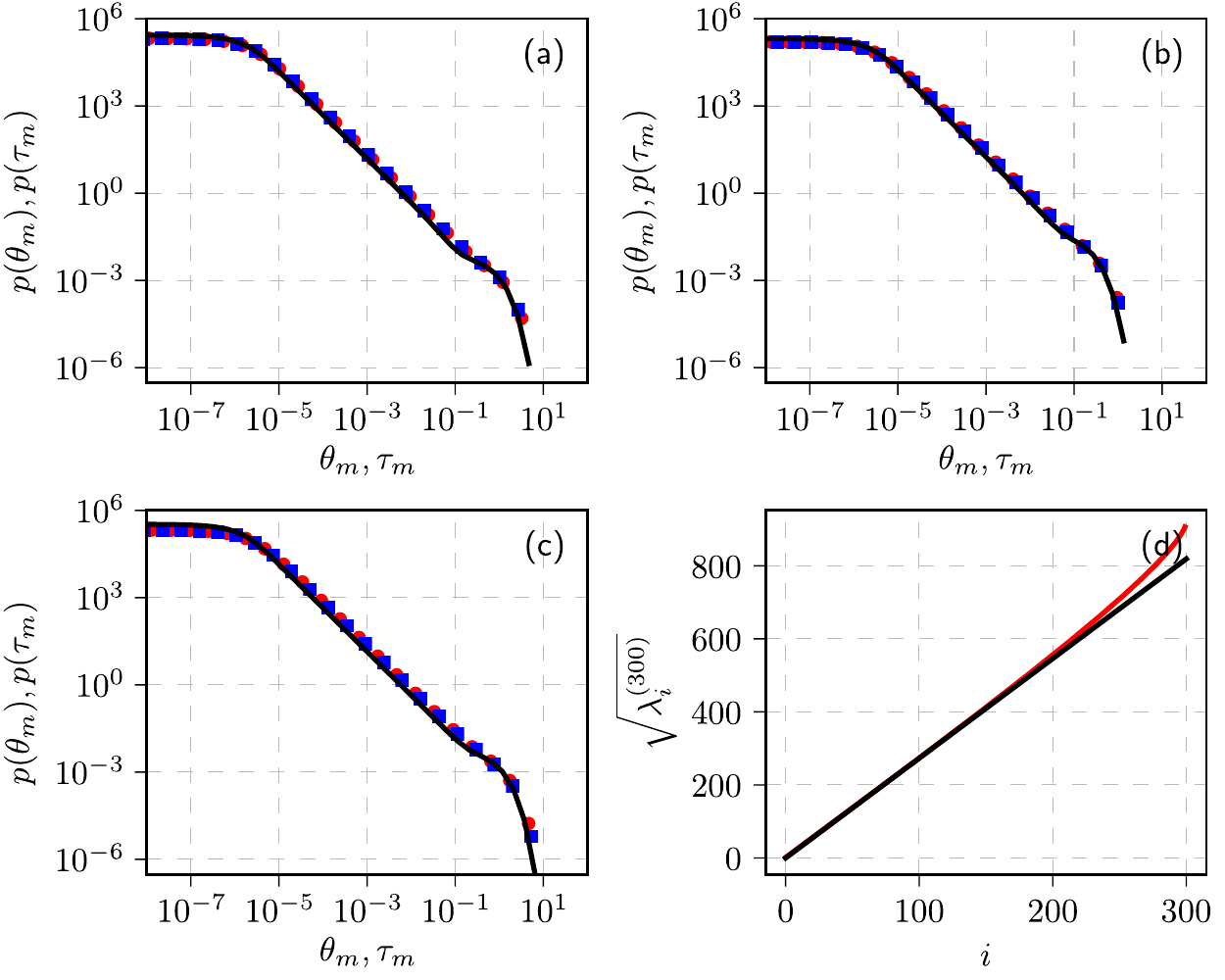}
\par\end{centering}
\caption{(color online) Burst and inter--burst duration PDFs ((a), (b) and
(c)) for the imitation process and square root of its eigenvalue spectrum
(sub--figure (d)). In (a), (b) and (c) red circles represent numerical
inter--burst duration PDF, blue squares represent numerical burst
duration PDF, while black curve was obtained by evaluating Eq.~(\ref{eq:bdApprox2}).
In (d) red curve represents a square roots of the eigenvalues, while
black curve show the best linear fit. Parameters of the process were
set as follows: (a) $\varepsilon=0.5$, $N=10^{3}$, $h=0.3$ (for
inter--burst duration) and $0.7$ (for burst duration), (b) $\varepsilon=1$,
$N=10^{3}$, $h=0.2$ (for inter--burst duration) and $0.8$ (for
burst duration), (c) $\varepsilon=1.5$, $N=10^{3}$, $h=0.7$ (for
inter--burst duration) and $0.3$ (for burst duration), (d) was obtained
with same parameters as (a).\label{fig:kirman}}

\end{figure}

In Fig.~\ref{fig:kirman} we show that burst and inter--burst durations
in this model are indeed equivalent in statistical sense, as blue
squares and red circles trace the same shape. Furthermore the approximation
by Eq.~(\ref{eq:bdApprox2}) provides rather good approximation for
that shape. Also in Fig.~\ref{fig:kirman} (d) we show that the square
root of the eigenvalue spectrum indeed grows linearly in a certain
region.

\section{Bursting statistics of the order book model exhibiting long--range
memory\label{sec:sophisticated-model}}

\begin{figure}
\begin{centering}
\includegraphics[width=0.7\textwidth]{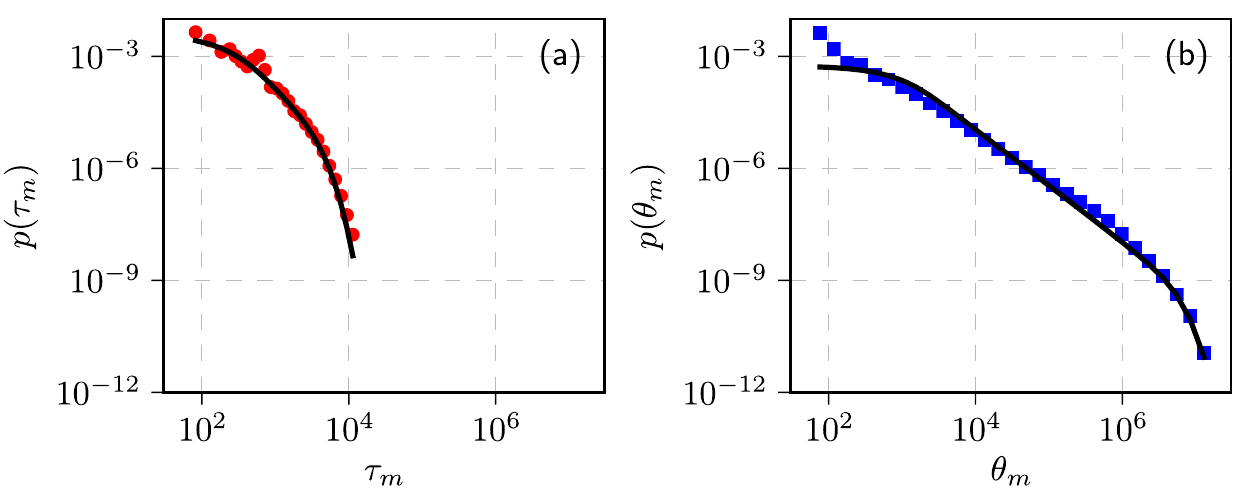}
\par\end{centering}
\caption{Burst (a) and inter--burst (b) duration PDFs of the numerical absolute
return time series. Numerical absolute return time series were obtained
by evaluating the order book model exhibiting long--range memory
using parameter values fitting the statistical properties of Bitcoin
time series best \cite{Kononovicius2019OB}. Threshold level was set
at $3$ standard deviations of the numerical time series. Numerical
burst duration PDF is given by red circles in (a) and numerical inter--burst
duration PDF is given by blue squares in (b), while black curves provide
the best fits according to Eq.~(\ref{eq:bdApprox2}). Parameters
of the fits were set as follows: (a) $\rho=0.15$, $\lambda_{1}=9.1\cdot10^{-4}$,
$\lambda_{2}=8.4\cdot10^{-3}$ and $\lambda_{m}=1.45\cdot10^{-3}$,
(b) $\rho=0.012$, $\lambda_{1}=5.1\cdot10^{-7}$, $\lambda_{2}=9.4\cdot10^{-7}$
and $\lambda_{m}=1.87\cdot10^{-3}$.\label{fig:arbitrary}}
\end{figure}

In this section we use Eq.~(\ref{eq:bdApprox2}) to fit bursting
statistics obtained from the order book model exhibiting long--range
memory \cite{Kononovicius2019OB}. From this model we have obtained
numerical absolute return time series. We have treated the series
using standard deviation filter with $10$ minute window as described
in \cite{Gontis2017PhysA}. By setting the threshold level at $3$
standard deviations we have obtained burst and inter--burst duration
PDFs, which are shown in Fig.~\ref{fig:arbitrary} as red circles
and blue squares.

As you can see in Fig.~\ref{fig:arbitrary} the fit provided by black
curves, which represent best fits according to Eq.~(\ref{eq:bdApprox2}),
is rather good. The parameters of the fits were obtained by matching
four empirical moments (mean, variance, third and fourth central moments)
against the same moments calculated for Eq.~(\ref{eq:bdApprox2}).
In both cases estimate of $\lambda_{m}$ was found to be reasonably
close to $\frac{1}{600}$, which coincides with the standard deviation
filter window we have used.

\section{Conclusions\label{sec:conclusions}}

Here we have proposed a method to approximate the first passage time
PDF between the successive states for any birth--death process. We
have shown that the method works on a three distinct birth--death
processes. As our derivation relies only on the general properties
of birth--death process and on the concept of Riemann sum, we believe
that the approximation should be applicable to any other birth--death
process. Furthermore first passage times are invariant to variable
transformations, thus the approximation method should be applicable
to any one--dimensional diffusive process, which can be transformed
into a birth--death process.

It is worth to note that our derivation suggests that all birth--death
processes should share a peculiar property. The square root of their
eigenvalue spectrum, $\sqrt{\lambda_{i}^{\left(n\right)}}$, should
grow linearly for a wide range of ranks, $i$. This property must
hold in order to be able to obtain power--law PDF with exponent $-3/2$,
which is a known result for all one--dimensional Markov processes
\cite{Redner2001Cambridge}.

We have also used the proposed method to fit burst and inter--burst
duration PDFs of the order book model exhibiting long--range memory
\cite{Kononovicius2019OB}. This model is based on a birth--death
process, but is significantly more sophisticated with other source
of randomness embedded into the model. While this model is not a one--dimensional
Markov process, but the approximation method seems to work reasonably
well. This example is also particularly interesting, because it is
able to reproduce power--law statistical properties of the Bitcoin
time series. Including the presence of long--range memory. The fact
that we are able to fit its burst and inter--burst duration PDFs
suggests that the considered order book model exhibits spurious long--range
memory, which could also be the nature of empirically observed long--range
memory phenomenon. Indeed one could use a similar fitting method to
fit the empirical burst and inter--burst duration PDFs. Yet to fully
undertake the empirical fitting one first needs to find reliable methods
to remove noise inherent to the empirical series.

\section*{Acknowledgement}

One of the authors acknowledges the funding provide by the European
Social Fund under the No 09.3.3--LMT--K--712 ``Development of Competences
of Scientists, other Researchers and Students through Practical Research
Activities'' measure.

%\bibliographystyle{ieeetr}
%\bibliography{library}

\end{document}